\documentclass[aps,pra,twocolumn,preprint,showpacs,preprintnumbers,amsmath,amssymb]{revtex4}

\usepackage{graphicx}
\usepackage{dcolumn}
\usepackage{bm}
\usepackage{color}
\usepackage{slashbox}
\usepackage{amsmath}

\newcommand{\nab}{{\bf{\nabla}}}

\newcommand{\lam}{\Lambda}

\newcommand{\pa}{\partial}

\newcommand{\larger}{{\bf r}}

\newcommand{\hanaV}{\mathcal{V}}

\begin{document}

\title{Rapid coherent control of population transfer in lattice systems}
\author{Shumpei Masuda$^{1,2}$}
\email[]{masuda@uchicago.edu}
\author{Stuart A. Rice$^{1}$}
\affiliation{
$^{1}$The James Franck Institute, The University of Chicago, Chicago, 
Illinois 60637, USA}
\affiliation{
$^{2}$Department of Physics,
Tohoku University,
Sendai 980, Japan}
\date{\today}
\begin{abstract}
We derive the driving potential that accelerates 
adiabatic population transfer from an initial state to 
a target state in a lattice system without unwanted excitation of 
other states by extending to discrete systems the fast-forward theory 
of adiabatic transfer.  
As an example we apply the theory to a model that describes 
a Bose-Einstein condensate in a quasi one-dimensional optical lattice, 
and show that modulation of the tilting of the lattice potential can 
transfer the population of the Bose-Einstein condensate 
from site to site with high fidelity and without unwanted excitations.
\end{abstract}
\maketitle

\section{Introduction}
During the last three decades there have been dramatic advances 
in both theoretical understanding of the requirements for control of 
quantum dynamics and the technology that is needed for the execution of 
proposed control paradigms\cite{ric2,Shap}.
Experimental verifications of the theory for systems as diverse as 
control of population transfer in Bose-Einstein condensates (BECs) and 
in chemical reactions have been reported 
\cite{Baum1,Gaub,Baum2,Ditt,Half,Scha,Scha2,Bas}. 
A particularly useful subgroup of the proposals for control of 
quantum dynamics of a system rely on adiabatic transfer via the slow 
variation of an external field that is applied to the system.  However, 
experimental exploitation of such control schemes can be rendered difficult 
by the occurrence of unwanted internal decoherence processes and by external 
noise; both of these difficulties can be reduced or avoided if the adiabatic 
transfer process can be speeded up sufficiently to permit population transfer 
to compete successfully with the time-dependence of the perturbations.  
Indeed, with this goal in 
mind, several methods for the acceleration of quantum dynamics, including 
adiabatic dynamics, have been proposed.  These methods include the 
counter-diabatic protocol \cite{ric}, frictionless quantum driving \cite{ber3}, 
invariant-based inverse engineering \cite{mug1}, 
and fast forward scaling \cite{mas1,mas2,mas3,mas4,torr2}, 
which is also used for protection of quantum 
states from potential ﬂuctuations \cite{mas5}.

Lattice models are widely used to describe quantum systems, examples of 
which are a BEC in an optical lattice, a network of nonlinear waveguides 
and optical fibers, and a superconducting ladder of Josephson junctions. 
For example, motivated by the potential applicability to quantum computation,
and by the opportunity to simulate aspects of complex electronic behavior in
crystalline matter, many remarkable features of BECs in optical lattices have
been studied \cite{Mor2}.
The existing studies clearly reveal the value of the ability to manipulate
BECs in optical lattices for the purpose of preparing well-defined quantum
states.
We have been stimulated by 
this observation to 
to extend 
the theory of accelerated adiabatic transfer to lattice systems so as to 
determine the potential that drives 
specified state-to-state population transfer 
without excitation of unwanted quantum states.  
In this paper we provide a 
derivation of that driving potential, and we apply the theory to site-to-site 
population transfer of a BEC in a quasi-one-dimensional 
optical lattice.  
We show that modulation of the lattice potential can 
transfer the population of the BEC between sites of the 
lattice with high fidelity and without unwanted excitations.  
The theory 
developed is applicable to any lattice in which the on-site potential is 
tunable.  We also demonstrate the robustness of the accelerated population 
transfer to variation (approximation) of the driving potential. 

In Sec. \ref{model} 
we present the framework of the theory of accelerated quantum 
adiabatic dynamics in a lattice system and discuss its relationship with 
the corresponding theory for a continuous system.  
In Sec. \ref{Control of BEC in optical lattice potential}  we study 
accelerated population transfer in a Bose-Einstein condensate in a 
one-dimensional optical lattice potential.  The robustness of the method 
with respect to approximation of the driving potential is studied 
in Sec. \ref{Comments}.  
An Appendix provides a brief description of the basic theory of acceleration 
of non-adiabatic quantum dynamics.

\section{Fast-forward transformation in discrete systems}
\label{model}
We consider a lattice system in which the dynamics is governed by a 
discrete time-dependent Schr$\ddot{\mbox{o}}$inger equation
\begin{eqnarray}
i\frac{d\Psi(m,t)}{dt} &=& \sum_l \tau_{m,l} \Psi(l,t) \nonumber\\
&&+ \frac{V_0(m,R(t))}{\hbar}\Psi(m,t),
\label{se0}
\end{eqnarray}
where $l,m$ denote sites and $t$ time, respectively, and
$\tau_{m,l} = \tau_{l,m}^\ast$ is the rate of hopping 
between sites $m$ and $l$.
The potential $V_0$ is modulated by a parameter $R$, which is a function of $t$.
If the parameter $R$ changes slowly enough from $R_i$  to $R_f$, 
and if the initial state is the $n$th energy eigenstate
of the Hamiltonian with potential $V_0(R_i)$, 
the wave function of the state on site $m$ 
changes from $\phi_n(m,R_i)$ to $\phi_n(m,R_f)$ modulo the 
dynamical and adiabatic phases of the states.  
The wave function $\phi_n(m,R)$ is a solution of the time-independent 
Schr$\ddot{\mbox{o}}$inger equation
\begin{eqnarray}
&&\sum_l \hbar\tau_{m,l} \phi_n(l,R) +  V_0(m,R)\phi_n(m,R) \nonumber\\
&&= E_n(R) \phi_n(m,R).
\label{se1}
\end{eqnarray}
On the other hand, when the parameter $R$ changes at a non-zero rate, 
transitions occur to other levels.  Our purpose is to derive a potential 
that drives the state from $\phi_n(m,R_i)$ to $\phi_n(m,R_f)$ 
in some short time $T_F$ without unwanted 
excitations to other states.  For that purpose we consider an intermediate 
state whose wave function is represented as
\begin{eqnarray}
\Psi_{FF}(m,t) &=& \phi_n(m,R(t))\exp\big{[}if(m,t)\big{]} \nonumber\\
&&\times\exp\Big{[}-\frac{i}{\hbar}\int_0^tE_n(R(t'))dt'\Big{]}.\ \ \ \
\label{psiFF}
\end{eqnarray}
Note that Eq. (\ref{psiFF}) contains the additional phase $f(m, t)$,
and that the intermediate state connects the initial state $\phi_n(m,R_i)$ and 
the target state $\phi_n(m,R_f)\exp\Big{[}
-\frac{i}{\hbar}\int_0^{T_F}E_n(R(t'))dt'\Big{]}$ in time $T_F$. 
We require that this additional phase 
vanishes at $t=0$ and at $t=T_F$, and 
we assume that the intermediate state satisfies the time-dependent
Schr$\ddot{\mbox{o}}$inger equation
\begin{eqnarray}
&&i\frac{d\Psi_{FF}(m,t)}{dt} = \sum_l \tau_{m,l} \Psi_{FF}(l,t) \nonumber\\
&&+  \frac{V_{FF}(m,t)}{\hbar}\Psi_{FF}(m,t),
\label{se2}
\end{eqnarray}
in which $V_{FF}(m,t)$ is the driving potential.
We seek the driving potential that generates
$\phi_n(m,R_f)\exp\Big{[}-\frac{i}{\hbar}\int_0^{T_F}E_n(R(t'))dt'\Big{]}$ 
from $\phi_n(m,R_i)$.
Although we do not aim to generate the adiabatic phase, that uniform 
phase can be tuned by a uniform potential if necessary. 

To find the forms of the driving potential and the additional phase $f(m,t)$
we substitute Eq. (\ref{psiFF}) into the Schr$\ddot{\mbox{o}}$inger 
equation (\ref{se2}) and we use Eq. (\ref{se1}) to rearrange the resulting 
equation.  The imaginary part of the resultant equation leads to
\begin{eqnarray}
&& \dot{R} \mbox{Re}\big\{ \phi_n^\ast(m,R)\pa_R\phi_n(m,R) \big\} \nonumber\\
&=& \sum_l \mbox{Im} \Big{(}
\tau_{m,l}\phi_n^\ast(m,R)\phi_n(l,R) \nonumber\\
&&\times \big\{ \exp\big{[} i\big{(}f(l,t) - f(m,t)\big{)} \big{]} - 1\big\} 
\Big{)}.\nonumber\\
\label{eq_f}
\end{eqnarray}
The solution of Eq. (\ref{eq_f}) yields the additional 
phase $f(m,t)$, and the real part gives the driving 
potential as a functional of $f$, $V_0$, $R$ and $\phi_n$: 
\begin{eqnarray}
&&V_{FF}(m,t) = V_0(m,R(t)) \nonumber\\
&&+ \sum_l \mbox{Re} \Big\{ \hbar\tau_{m,l}
\frac{\phi_n(l,R(t))}{\phi_n(m,R(t))} \nonumber\\
&& \times \big{(}1-\exp\big{[}i\{f(l,t)-f(m,t)\}\big{]}\big{)} \Big\} 
\nonumber\\
&&-\hbar\dot{f}(m,t)
- \hbar\dot{R}
\mbox{Im}\Big{[}\frac{\pa_R\phi_n(m,R(t))}{\phi_n(m,R(t))}\Big{]}.
\label{eq_VFF}
\end{eqnarray}
It is necessary that
$R$ satisfies the conditions
\begin{eqnarray}
R(0) &=& R_i\nonumber\\
R(T_F) &=& R_f.
\end{eqnarray}
If we take the boundary conditions to be
\begin{eqnarray}
\dot{R}(0) = \dot{R}(T_F) = 0,
\end{eqnarray}
$f(m,t)$ vanishes at $t=0$ and at $t=T_F$ 
(see Eq. (\ref{eq_f}) ), 
and the intermediate state coincides with the target state at $T_F$. 
The driving potential is obtained by substituting the additional 
phase into Eq. (\ref{eq_VFF}). 
With the boundary conditions
\begin{eqnarray}
\ddot{R}(0) = \ddot{R}(T_F) = 0
\end{eqnarray}
the driving potential coincides with $V_0$ at $t=0$ and at $t=T_F$.
The time-dependence of R is arbitrary except for the requirement 
imposed by the above boundary conditions.  The driving potential 
depends on the time-dependence of $R$. 

In the case that the hopping rate and the wave function are real, 
Eq. (\ref{eq_f}) and Eq. (\ref{eq_VFF}) simplify to
\begin{eqnarray}
&&\ \dot{R} \pa_R\phi_n(m,R) \nonumber\\
&=& \sum_l\tau_{m,l}\phi_n(l,R) \sin[ f(l,t) - f(m,t) ],\nonumber\\
\label{eq_f2}
\end{eqnarray}
and
\begin{eqnarray}
&&V_{FF}(m,t) = V_0(m,R(t)) \nonumber\\
&&+ \sum_l \hbar\tau_{m,l}
\frac{\phi_n(l,R(t))}{\phi_n(m,R(t))} \nonumber\\
&& \times \{1-\cos[f(l,t)-f(m,t)]\} - \hbar\dot{f}(m,t).\nonumber\\
\label{eq_VFF2}
\end{eqnarray}


We note that Eq. (\ref{eq_f})  implies that for $\dot{R}$
sufficiently large there is no solution for $f(m,t)$.  
That is, there is a lower limit to the control time $T_F$.  
This property is not seen in the fast-forward theory for continuous 
systems \cite{mas2}.  
Eqs. (\ref{eq_f}) and (\ref{eq_VFF}), for $f$ and for $V_{FF}$, 
reduce to the corresponding equations for continuous systems 
shown in Ref. \cite{mas2} in the limit that the diﬀerences between 
adjacent sites of $f$ and of $\phi_n$ are small. 
The theory of acceleration of non-adiabatic quantum dynamics in a
continuous system is described in Ref. \cite{mas1}.
Following the same analysis as in Ref. \cite{mas1}, 
the key elements of the theory of accelerated non-adiabatic 
quantum dynamics in a lattice system are exhibited in the Appendix.

The analysis described above can be 
straightforwardly extended to the case when a nonlinear 
Schr$\ddot{\mbox{o}}$inger equation is the basic descriptor of the 
system dynamics.  
Consider 
\begin{eqnarray}
&&i\frac{d\Psi(m,t)}{dt} = \sum_l \tau_{m,l} \Psi(l,t) 
\nonumber\\
&&+ \frac{V_0(m,R(t))}{\hbar}\Psi(m,t) + \frac{c}{\hbar}|\Psi(m,t)|^2\Psi(m,t),
\nonumber\\
\label{se0_non}
\end{eqnarray}
where $c$ is a constant.
We assume the same form of the wave function of the intermediate state 
$\Psi_{FF}$ as in Eq. (\ref{psiFF}).  
Then $\phi_n$ is a solution of the time-independent nonlinear 
Schr$\ddot{\mbox{o}}$inger equation 
\begin{eqnarray}
&&\sum_l \hbar\tau_{m,l} \phi_n(l,R) +  V_0(m,R)\phi_n(m,R) \nonumber\\
&&+ c|\phi_n(m,R)|^2\phi_n(m,R)= E_n(R) \phi_n(m,R).\nonumber\\
\label{se1_non}
\end{eqnarray}
We assume that the intermediate state wave function 
is defined by the nonlinear Schr$\ddot{\mbox{o}}$inger equation 
\begin{eqnarray}
&&i\frac{d\Psi_{FF}(m,t)}{dt} = \sum_l \tau_{m,l} \Psi_{FF}(l,t) 
\nonumber\\ 
&&\hspace{2.3cm} +  \frac{V_{FF}}{\hbar}(m,t)\Psi_{FF}(m,t) \nonumber\\ 
&&\hspace{2.3cm} + \frac{c}{\hbar}|\Psi_{FF}(m,t)|^2\Psi_{FF}(m,t).\ \ \ \ \
\label{se2_non}
\end{eqnarray}
We can derive the equations for the additional phase and the 
driving potential in the same manner as for the 
linear Schr$\ddot{\mbox{o}}$inger equation.  
The resultant equations are the same as 
Eqs. (\ref{eq_f}) and (\ref{eq_VFF}), respectively.  
The nonlinear term influences the driving potential through $\phi_n$ 
in Eq. (\ref{se1_non}).

\section{Site-to-site population transfer of a BEC in an optical lattice}
\label{Control of BEC in optical lattice potential}
As an example, we now consider site-to-site population transfer of 
a BEC in an optical lattice.  The lattice is defined by an external potential 
that is the sum of a spatially linear potential, which is tunable, 
and a stationary periodic potential
\begin{eqnarray}
V_{ext}(\larger,t) = \xi(t) z + U_L(x,y) \sin^2 (2\pi z/\lambda),\ 
\label{Vext1}
\end{eqnarray}
where $\lambda/2$ is the wavelength (period) of the potential.
We consider the case that the mean ﬁeld condensate interaction is negligible.  
A discrete model of the BEC in a tilted trap was introduced in Ref. \cite{Tro}, 
using the tight binding approximation.  
In the tight binding approximation the condensate order parameter is written as
\begin{eqnarray}
\Phi(\larger,t) = \sqrt{N_T}\sum_m\Psi(m,t)\varphi(\larger-\larger_m),
\label{eq_Phi}
\end{eqnarray}
where $N_T$ is the total number of atoms and 
$\varphi(m,\larger)=\varphi(\larger-\larger_m)$
is the condensate wave function localized in the $m$th trap with location
$\larger_m$.
We assume that
$\int \varphi(m,\larger)\varphi(m+1,\larger)d\larger = 0$ 
and $\int \varphi^2(m,\larger) d\larger = 1$.
Using Eq. (\ref{eq_Phi}),
the Gross-Pitaevskii equation can be rewritten to read \cite{Tro}
\begin{eqnarray}
i\hbar\frac{\pa}{\pa t}\Psi(m,t) &=& 
- K\big{[}\Psi(m-1,t)+\Psi(m+1,t)\big{]}\nonumber\\
&&+ \frac{\xi(t)\lambda m}{2} \Psi(m,t),
\label{GP1}
\end{eqnarray}
where 
\begin{eqnarray}
K &\simeq& -\int d\larger\Big{[} 
\frac{\hbar^2}{2m_0}\nab\varphi(m,\larger)\cdot\nab \varphi(m+1,\larger)
\nonumber\\
&&+ \varphi(m,\larger) V_{ext}(\larger) \varphi(m+1,\larger)\Big{]},
\end{eqnarray}
with $m_0$ the mass of an atom.
$K$ is independent of $m$ because of the orthogonality
$\int \varphi(m,\larger)\varphi(m+1,\larger)d\larger = 0$.  
Equation (\ref{GP1}) then can be rewritten as 
\begin{eqnarray}
i\frac{\pa}{\pa t}\Psi(m,t) &=& 
\tau\big{[}\Psi(m-1,t)+\Psi(m+1,t)\big{]}\nonumber\\
&&+ \frac{V(m,t)}{\hbar} \Psi(m,t),
\label{GP2}
\end{eqnarray}
with 
\begin{eqnarray}
\tau &=& -K/\hbar,
\end{eqnarray}
and
\begin{eqnarray}
V(m,t) &=& \frac{1}{2}\xi(t)\lambda m.
\end{eqnarray}
We demonstrate the acceleration of population transfer 
for a BEC in a lattice with this model.  
Our goal is the transfer of population to the ground state of 
the linear potential with $\xi=\xi_f$ from the ground state of the 
linear potential with $\xi=\xi_i$.
We take $\xi_i = -\xi_f$ 
so that the population is transferred from one side of the lattice 
to the opposite side of the lattice. 

\subsection{A three-site model}
\begin{figure}[h!]
\begin{center}
\includegraphics[width=7.5cm]{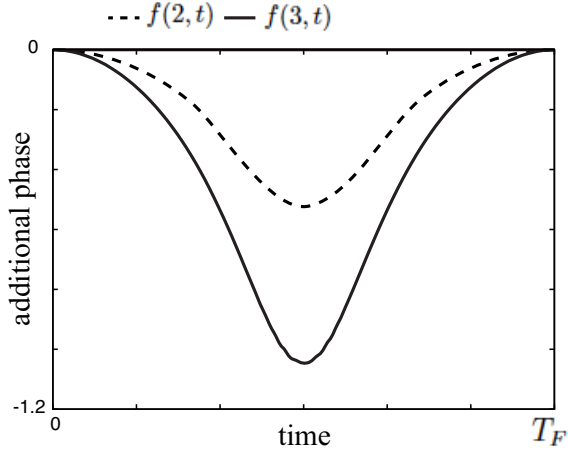}
\end{center}
\caption{\label{fig:epsart}  
Time-dependence of the additional phase.}
\label{f_stf3_3site}
\end{figure}

\begin{figure}[h!]
\begin{center}
\includegraphics[width=7.5cm]{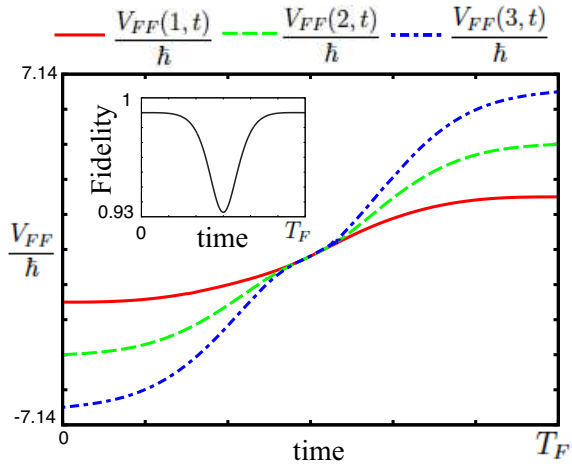}
\end{center}
\caption{\label{fig:epsart} (Color online) 
Time-dependence of $V_{FF}(m,t)/\hbar$.
The unit of time is 1 ms.
The inset shows the time-dependence of the fidelity, defined
by $F(t) = |<\phi_0|\Psi>|$.}
\label{VFF_stf3}
\end{figure}

\begin{figure}[h!]
\begin{center}
\includegraphics[width=7.5cm]{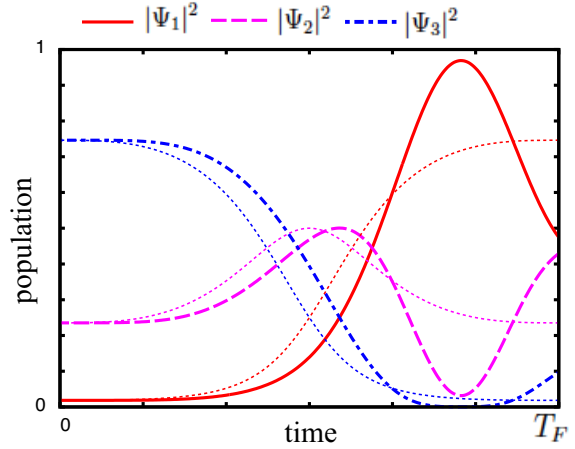}
\end{center}
\caption{\label{fig:epsart} (Color online) 
Time-dependence of the population evolution under $V_0(m,R(t))$
(dashed and solid lines)
and $V_{FF}(m,t)$ (dotted lines).
The evolution under the instantaneous Hamiltonian is also shown 
with dotted lines.  
The notation is $\Psi_m=\Psi(m,t)$. }
\label{dyn_H0_stf3}
\end{figure}

We consider first a three-site model with site potential
\begin{eqnarray}
V_0(m,R(t)) =  \hbar\omega R(t) m.
\label{V01}
\end{eqnarray}
In Eq. (\ref{V01}), the constant frequency $\omega$ is defined by
\begin{eqnarray}
\omega = -\frac{\xi_i\lambda}{2\hbar} = \frac{\xi_f\lambda}{2\hbar},
\end{eqnarray}
and the time-dependence of $R(t)$ is chosen to be
\begin{eqnarray}
R(t) = R_0 + \frac{2}{T_F}\Big{[}
t-\frac{T_F}{2\pi}\sin\Big{(}\frac{2\pi}{T_F}t\Big{)} \Big{]}.
\end{eqnarray}
We take $R_0=-1$,
so that $V_0(m,R(t))$ changes from $\xi_i\lambda m /2$
to $\xi_f\lambda m /2$ in time $T_F$, and take
the hopping rate in Eq. (\ref{se0}) to be
\begin{eqnarray}
\tau_{m,l} = \tau (\delta_{m,l-1} + \delta_{m,l+1}).
\end{eqnarray}
We calculated the additional phase and driving potential for this 
model system using Eqs. (\ref{eq_f2}) and (\ref{eq_VFF2}), respectively, 
with the parameter set $T_F=4.2$ ms, $\omega=2.14$ /ms,
$\hbar/2K=0.35$ ms and $\lambda=850$ nm \cite{Tro}.
The time-dependence of the additional phase is shown in Fig. \ref{f_stf3_3site},
where we choose $f(1,t)=0$. 
The driving potential $V_{FF}(m,t)$, shown in Fig. \ref{VFF_stf3},
differs from $V_0(m,R(t))$ for $0<t<T_F$, 
and is equal to $V_0$ at $t=0$ and $t=T_F$.
We have simulated the evolution of the model system driven by $V_{FF}(m,t)$ 
from the ground state corresponding to $V_{0}(m,R(0))$.  
That evolution is monitored by the fidelity
\begin{eqnarray}
F(t) = |<\phi_0|\Psi>|,
\end{eqnarray}
where $|\phi_0>$ is the ground state of the instantaneous 
Hamiltonian $H_0(R(t))$ and $|\Psi>$ is 
the state driven by the potential $V_{FF}(m,t)$.
The time-dependence of the fidelity is shown in the inset to 
Fig. \ref{VFF_stf3}; 
it is equal to unity at $T_F$.  
A comparison of the population evolution under $V_0(m,R(t))$ and 
under $V_{FF}(m,t)$ is shown 
in Fig. \ref{dyn_H0_stf3}.  
We note that the non-adiabatic transfer generates unwanted 
excitations, with the population of each site deviating from that 
evolving under the instantaneous Hamiltonian 
(dotted lines in Fig. \ref{dyn_H0_stf3}).  
The fidelity of the population evolution driven by $V_0(m,R(t))$ 
is 0.938 at $T_F$.
\subsection{A four-site model}
\begin{figure}[h!]
\begin{center}
\includegraphics[width=7.5cm]{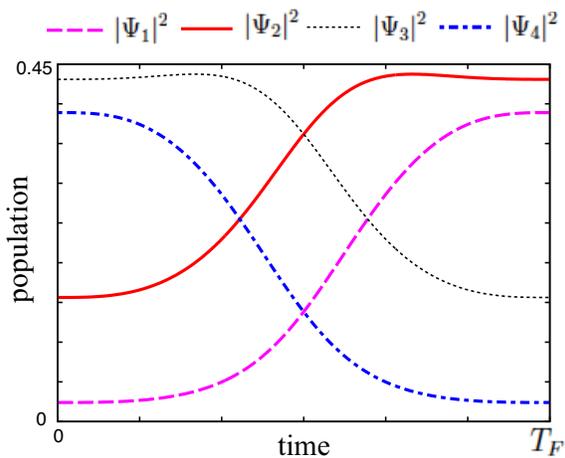}
\end{center}
\caption{\label{fig:epsart} (Color online) 
Population of the ground state of the instantaneous Hamiltonian 
in the four-site model system.}
\label{p_ex_4site}
\end{figure}

\begin{figure}[h!]
\begin{center}
\includegraphics[width=7.5cm]{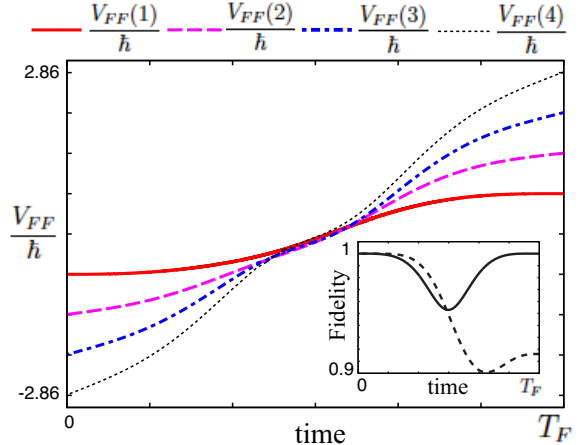}
\end{center}
\caption{\label{fig:epsart} (Color online) 
Time-dependence of $V_{FF}(i,t)/\hbar$.
The unit of time is 1 ms.
The inset shows the time-dependence of the fidelity.}
\label{VFF_4site}
\end{figure}

We have also examined accelerated population transfer of a BEC in a 
four-site model.  
The parameters used for these calculations are the same 
as for the three-site model except that $\omega = 0.714$ /ms.  
The population of the ground state of the instantaneous Hamiltonian 
for each site is shown in Fig. \ref{p_ex_4site}.
The initial state is located mainly at sites 3 and 4, 
while the target state is located mainly at sites 1 and 2.  
The time-dependence of the driving potential is shown in Fig. \ref{VFF_4site}.  
The time-dependence of the fidelity are compared 
in the inset to Fig. \ref{VFF_4site}.  
The solid curve and the broken curve correspond to the 
dynamics with $V_{FF}$ and $V_0$, respectively.  
We note that the fidelity 
decreases and does not recover at $T_F$ in the $V_0$ generated dynamics 
because of unwanted excitations whereas for the $V_{FF}$ generated dynamics 
the fidelity becomes unity at $t = T_F$.  

\section{Comments}
\label{Comments}
\begin{figure}[h!]
\begin{center}
\includegraphics[width=7.5cm]{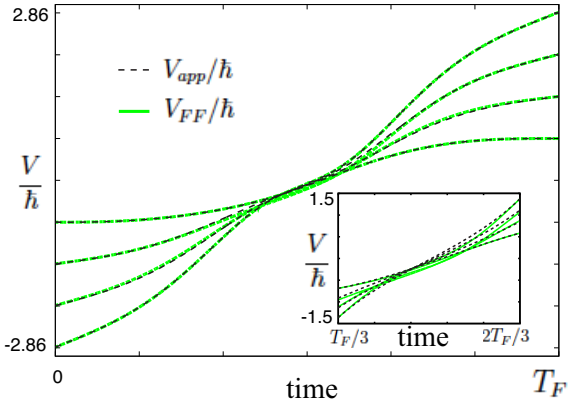}
\end{center}
\caption{\label{fig:epsart} (Color online) 
Comparison of $V_{app}/\hbar$ and $V_{FF}/\hbar$.
The unit of the vertical axis is 1/ms.
The inset shows the time-dependence of $V/\hbar$ for $T_F/3\le t\le 
2T_F/3$.}
\label{VFF_com_4site}
\end{figure}

\begin{figure}[h!]
\begin{center}
\includegraphics[width=7.5cm]{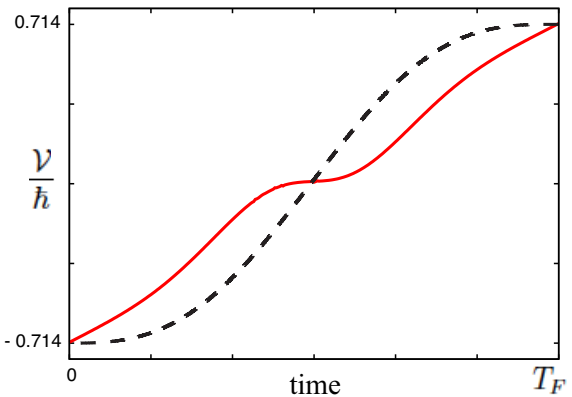}
\end{center}
\caption{\label{fig:epsart} (Color online) 
Comparison of $\hanaV (t)/\hbar$ (red solid curve) and $\omega R(t)$ (black
broken curve).
The unit of the vertical axis is 1/ms.}
\label{V0_VFF_com_4site}
\end{figure}

It is one matter to calculate the exact driving potential required 
to transfer the BEC population between sites with perfect fidelity, 
but it is another matter to generate that potential in a real experiment.  
It is usually the case that in real experiments we cannot generate a 
perfect rendition of a specified potential.  
Then, the robustness of the proposed population transfer method to 
variation of the driving potential is important.   
We can test the efficiency of our proposed transfer process 
to approximation of the driving potential by considering population 
transfer under a driving potential that is proportional to the site number:
\begin{eqnarray}
V_{app}(j,t) = \hanaV(t) j.
\label{eq_Vapp}
\end{eqnarray}
In Eq. (\ref{eq_Vapp}), $\hanaV(t)$ 
is a function designed so that $V_{app}$ approximates
the exact driving potential.
For the three-site model, for transfers between ground states, 
$V_{app}$ coincides with $V_{FF}$ because
\begin{eqnarray}
&&\phi_n(1,R) \big{[}2\phi_n^2(3,R)-\phi_n^2(2,R)\big{]} \nonumber\\
&& \ = \phi_n(3,R) \big{[}2\phi_n^2(1,R)-\phi_n^2(2,R)\big{]},\ \ \
\end{eqnarray}
for any $R$.
This property also holds for second and third eigenstates of the
instantaneous Hamiltonian, although the driving potential 
depends on the level $n$. 
Thus the simple potential defined in Eq. (\ref{eq_Vapp}) can 
transfer population in the three-site model without unwanted excitation.
The approximation $V_{app}(j,t) = \hanaV(t)j$ 
is not exact for the four-site model, 
but it is a good approximation to $V_{FF}$ for that model.  We show the 
difference between $V_{app}$ and $V_{FF}$ for the four-site model in 
Fig. \ref{VFF_com_4site}. 
In general, $V_{FF}$ is well approximated by $V_{app}$ with a larger deviation 
near $t = T_F /2$ than in other time domains (Fig. \ref{V0_VFF_com_4site}). 
The fidelity of 
the population transfer in the four-site system driven by $V_{app}$ is 
0.997 at $T_F$ whilst the fidelity of the population transfer 
driven by $V_0$ is 0.916.

Our derivation of the driving potential that accelerates adiabatic 
population transfer in a lattice reveals a striking difference between 
a lattice system and a continuous system.  
Specifically, in the lattice system there is lower limit to $T_F$.  
This limit derives from the condition for the additional phase in 
Eq. (\ref{eq_f}), 
which gives the lower limit for $\dot{R}$ for each $R$ 
depending on $\phi_n(R)$, that is, 
trajectory of the evolution of the system.  
We believe that the accelerated population transfer scheme described 
in this paper can be used for the coherent control of many quantum 
systems which are described by chain or lattice models.

\appendix

\section{Acceleration of non-adiabatic dynamics}
\label{Regular fast-forward}
We consider the acceleration of non-adiabatic quantum dynamics.
Consider the wave function $\Psi(m,t)$, which is a solution of 
a discrete time-dependent Schr$\ddot{\mbox{o}}$inger equation:
\begin{eqnarray}
i\frac{d\Psi(m,t)}{dt} &=& \sum_l \tau_{m,l} \Psi(l,t)\nonumber\\ 
&&+  \frac{V(m,t)}{\hbar}\Psi(m,t).
\end{eqnarray}
We seek a driving potential that generates the target state
$\Psi(m,T)$ at $t=T_F (< T)$.
We assume that the wave function of the intermediate state is 
\begin{eqnarray}
\Psi_{FF}(m,t) = \Psi(m,\lam(t))e^{if(m,t)},
\end{eqnarray}
where $f(m,t)$ is the additional phase and 
\begin{eqnarray}
\lam(t) = \int_0^t \alpha(t')dt'.
\end{eqnarray}
$\alpha$ is a real function of time called magnification factor \cite{mas1}.
The time-dependence of $\alpha$ is chosen so that it satisfies
\begin{eqnarray}
\lam(T_F) = T.
\end{eqnarray}
We assume that $\Psi_{FF}(m,t)$ is a solution of the 
Schr$\ddot{\mbox{o}}$inger equation:
\begin{eqnarray}
&&i\frac{d\Psi_{FF}(m,t)}{dt} = \sum_l \tau_{m,l} \Psi_{FF}(m,t) \nonumber\\
&&\hspace{1cm} +  \frac{V_{FF}(m,t)}{\hbar}\Psi_{FF}(m,t),
\end{eqnarray}
where $V_{FF}$ is the driving potential.
Following the same analysis as in Sec. \ref{model} we find
\begin{eqnarray}
&&\alpha(t)\sum_l  \mbox{Im} [\tau_{m,l} \Psi_m^\ast\Psi_l]\nonumber\\
&&= \sum_l \mbox{Im}\Big\{\tau_{m,l} \Psi_m^\ast \Psi_l 
\exp\big{[}i(f_l-f_m)\big{]}\Big\},\
\nonumber \\
\label{eq_f3}
\end{eqnarray}
and
\begin{eqnarray}
&&V_{FF}(m,t) = \nonumber\\
&&\sum_l \mbox{Re}\Big\{
\frac{\hbar\tau_{m,l}\Psi_l}{\Psi_m}\Big{[}\alpha(t)-e^{i(f_l-f_m)} \Big{]}
\Big\}\nonumber\\
&&+ \alpha(t)V(m,\lam(t)) - \hbar\pa_t f_m,\
\label{eq_VFF3}
\end{eqnarray}
where $f_m$ and $\Psi_m$ are abbreviations for
$f(m,t)$ and $\Psi(m,\Lambda(t))$, respectively.
Equation (\ref{eq_f3}) is used to obtain the additional phase.
The driving potential is obtained by substitution of $f_m$ 
into Eq. (\ref{eq_VFF3}).
As in the case of acceleration of adiabatic population transfer 
there is a lower limit to $T_F$ because Eq. (\ref{eq_f3}) gives the
upper limit of $\alpha(t)$ for each $t$.
The equations for $f$ and $V_{FF}$ in Eqs. (\ref{eq_f3}) and (\ref{eq_VFF3})
reduce to those for continuous systems in Ref. \cite{mas1}
in the limit that the differences in $f(m,t)$ and $\Psi(m,t)$ 
between adjacent sites are small.

\begin{acknowledgments}
SM thanks Grants-in-Aid for Centric Research of 
Japan Society for Promotion of Science
and JSPS Postdoctoral Fellowships for Research Abroad
for its financial support.
\end{acknowledgments}

\end{document}